# Local and electronic structure of $Sr_{1-x}Gd_xTiO_3$ probed by X-ray absorption spectroscopy

Alexandre Mesquita[1], Elio Thizay Magnavita Oliveira[2], Hugo Bonette de Carvalho[2]

[1] *Institute of Geosciences and Exact Sciences, São Paulo State University – Unesp, Rio Claro, SP, Brazil.*

[2] *Federal University of Alfenas – Unifal-MG, Alfenas, MG, Brazil.*

**Abstract**

Gadolinium-doped strontium titanate ($Sr_{1-x}Gd_xTiO_3$) is a typical perovskite structure material which has been studied due their thermomechanical, termoelectrical and electrochemical properties. In this study, local and electronic structure of $Sr_{1-x}Gd_xTiO_3$ samples were analyzed through X-ray absorption spectroscopy measurements. The results obtained with the adjustment of extended X-ray absorption fine structure (EXAFS) spectra at Sr K-edge show that crystallographic model of *Pm-3m* space group is consistent with local structure around Sr atoms, as expected. This same analysis also reveals an increasing of the Debye-Waller as a function of the Gd content in some shells, which is associated with disorder induced by Sr vacancies due to the heterovalent Gd incorporation. EXAFS spectra at Gd $L_{III}$-edge for $Sr_{1-x}Gd_xTiO_3$ samples indicates regular $GdO_{12}$ dodecahedra without displacement of Gd atoms from centrosymmetric position. A disorder was also identified in the shells beyond the first 12 O neighbors in which neither the crystallographic cubic structure of the $SrTiO_3$ nor the orthorhombic structure of the $GdTiO_3$ fits well. X-ray absorption near edge spectroscopy (XANES) spectrum at Ti $L_{III,II}$-edges shows an asymmetric peak because of the splitting between the $e_g$ orbitals of $3d$ band for $SrTiO_3$ sample. The addition of Gd atoms to $SrTiO_3$ structure cause an enlargement of this peak and this split is associated with a small displacement of Ti atoms from their centrosymmetric position. Several features of the XANES spectra at O k-edge for $Sr_{1-x}Gd_xTiO_3$ samples are affected by the increase of Gd concentration. According to our calculated projected density of states, these transitions are related to a reduction in the number of unoccupied O $2p$ - Ti $3d$ states caused by the split of Ti $3d$ band. Moreover, these XANES spectra also show a dependence of the increasing of the hybridization between O $2p$ and Gd $5d4f6s$ states.

## 1. Introduction:

Perovskite oxides ($ABO_3$) are a versatile and multifunctional class of material that has been studied extensively due to its potential applicability in a broad scope of fundamental and technological areas [1]. Among the perovskite oxides, $SrTiO_3$ (STO) is one of the most prominent material, it is a quantum paraelectric *n*-type large bandgap semiconductor (~ 3.3 eV) [2] with high dielectric permittivity (ε) and low dielectric loss (tan δ) [3], presenting very interesting properties under specific circumstances such as superconductivity [4] and ferroelectricity [5]. The STO has been used in varistors [6], capacitors [7], and studied also as based material in solar cell [8], thermoelectric [9], resistive-switching memory [10], piezoelectric [11], and gas sensing devices [12], just to cite few.

Among the strategies to enhance and add new functionalities to the STO, defect engineering has become an effective and powerful tool. In such a context, the structural doping is one of the most often used techniques [13–15]. Specifically, donor trivalent rare earth ($RE^{3+}$) substitutional doping of the perovskite STO structure at the $Sr^{2+}$ site, according to the stoichiometry $Sr_{1-x}RE_xTiO_3$, can substantially improve its thermoelectrical and [16,17] electrical properties [18,19]. Nevertheless, there are only few reports on the effect of Gd-doping over the STO properties. Since the $Gd^{3+}$ substitutional doping of the $Sr^{2+}$ increases effectively the electron concentration, it will lead to relatively larger values for the electron conductivity. Besides, due to the larger mass of Gd than that of Sr, Gd-doping would effectively lower the thermal conductivity of the STO due to a larger mass fluctuation effect that enhances the phonon scattering processes [20], corresponding to a higher thermoelectrical performance [21]. It was also shown that the proper amount of Gd-doping leads to an insulator-metal transition in STO [22], what would allow, for instance, its application as an electromagnetic interference shielding [23]. Even more, once the $Gd^{3+}$ crystal radius (1.41 Å) is smaller as compared to that for the $Sr^{2+}$ (1.58 Å), the $Gd^{3+}$ will occupy off-center positions in the STO structure, inducing an elastic lattice strain that suppress the intrinsic quantum fluctuation that hinders the observation of any ferroelectric phase transition [24,25].

In this scenario, we present in this report an X-ray absorption spectroscopy (XAS) study concerning several atom edges. This analysis aimed to understand how the local structure and the hybridization states of $Sr_{1-x}Gd_xTiO_3$ (SGTO) ceramic compounds are affected by the substitution of $Sr^{2+}$ by $Gd^{3+}$ ions. Despite several studies dedicated to this composition, the short-range order structure and electronic properties of SGTO system studied using XAS has not been found in the literature. XAS has been an important characterization technique to describe the local and medium-range atomic structure. This technique provides information on the electronic and structural properties around the absorber element choosing the proper absorption edge, which enable the use of this tool in several $ABO_3$ systems in amorphous and crystalline solids, dispersed systems and thin films [26]. Furthermore, the information obtained via this technique has been usefully correlated with thermoelectric properties [27–29]. In this paper, the local and electronic structures of SGTO powders were probed by measurements of the XAS spectra at the Ti and Gd $L$-edges and the Sr and O $K$-edges. XAS spectra in extended X-ray absorption fine structure (EXAFS) region were applied to evaluate the local structure of these atoms. *Ab initio* FEFF9 code calculation [30] were also used to simulate the obtained X-ray absorption near edge structure (XANES) spectra via hybridization states.

1. **Experimental procedure**

Crystalline $Sr_{1-x}Gd_xTiO_3$ (SGTO) ceramic samples with $x$ = 0.0, 0.06, 0.12, 0.18, 0.24 and 0.30 (0, 6, 12, 18, 24 and 30 at.%) were prepared by the polymeric precursor method (samples were labeled as SGTO$x$) [31]. Strontium nitrate $Sr(NO_2)_2$ (98%, Aldrich), titanium isopropoxide $Ti[OCH(CH_3)_2]_4$ (97%, Aldrich), gadolinium oxide $Gd_2O_3$ (99.9%, Aldrich), ethylene glycol $C_2H_6O_2$ (99.5%, Aldrich) and citric acid $C_6H_8O_7$ (99%, Aldrich) were used as precursors. The dissolution of the titanium isopropoxide in citric acid (CA) aqueous solution at constant stirring in 70 °C was performed in order to form the titanium citrate. Stoichiometric quantities of strontium nitrate, and gadolinium oxide $Gd_2O_3$ (dissolved in hot citric acid) were added to the Ti citrate solution. After homogenization, the temperature of the solution was raised up to around 110 °C and ethylene glycol (EG) was added. Then, the polyesterification reaction was promoted. The CA/metal and CA/EG mass ratios were fixed at 4/1 and 60/40, respectively. The samples were then heated at 300 °C for 2 h to eliminate the organic components. After pyrolysis, all samples were thermal treated at 700 °C for 2 h.

X-ray absorption near edge spectroscopy (XANES) and extended X-ray absorption fine structure (EXAFS) measurements were collected at LNLS (National Synchrotron Light Laboratory) facility. The measurements at Sr K- and Gd $L_{III}$-edges (16105 and 7243 eV, respectively) of SGTO samples were collected in transmission mode at

XAFS2 beam line. XANES spectra at Ti $L_{III,II}$- and O K-edges were collected at the PGM beamline. These XANES spectra (around 453 and 543 eV, respectively) were measured at room temperature with electron-yield mode. The multi-platform applications for X-ray absorption (MAX) software package [32] was used to extract and to fit the EXAFS spectra. Theoretical spectra (XANES and EXAFS) were obtained using the FEFF9 code [30]. This code was also used to obtain the projected density of states (DOS) [33].

2. **Results and discussions**

Figure 1(a) presents the XANES spectra at Sr K-edge for SGTO samples. A XANES spectra can provide the valence of the atoms in a material as well as its coordination symmetry. Depending on the valence of the absorbing ion, a shift in the position of the absorbing edge can be detected because of the binding energy of bound electrons which modifies with increasing of the valence. Moreover, the coordination symmetry and the unfilled local DOS have influence on the shape of the absorption edge of the absorbing element. As expected in the $SrTiO_3$ structure, the XANES spectra observed in Fig. 1(a) are consistent with $Sr^{2+}$ valence [34]. No significant modifications as a function of the Gd concentration are observed except for a small decrease in intensity of the transitions at around 16150 and 16200 eV (Insets of Fig. 1(a)).

In order to comprehend the origin of the oscillations observed in the post-edge region of these spectra, we performed successfully *ab initio* calculations of theoretical XANES spectra at Sr K-edge as a function of the cluster radius. The calculation was performed through *ab initio* FEFF code [30] with cluster radius of 6.0, 4.77, 3.91 and 3.38 Å and crystallographic model according to cubic $SrTiO_3$ structure with *Pm-3m* space group [35]. Figure 1(b) presents the obtained theoretical spectra. First we call attention that, by comparison between Fig. 1(a) and (b), the theoretical results reproduces satisfactorily the experimental obtained spectra. Inserted in Fig. 1(b) we see the illustration of each used cluster. The cluster with radius of 3.38 Å involves the first shell with 12 O atoms and the second shell with 8 Ti atoms around the absorber Sr atom, whereas the third shell with 6 Sr atoms is added in the cluster with radius equal to 3.91 Å. In addition to these 26 neighbors, the cluster with radius equal to 4.77 Å encompasses a shell with 24 O atoms. Finally, a shell with 11 Sr atoms is inserted, resulting in the cluster with radius equal to 6.00 Å and totaling 61 atoms around the absorber atom. We observe that there are no significant differences between the theoretical spectra for the 6.00 and 4.77 Å. As the cluster size is reduced from 4.77 to 3.91 Å, the main features are still observed, however the intensity of the peaks between 16130 and 16170 eV (Inset of Fig. 1(b)) decreases. As previously mentioned, the difference between these clusters is a shell of 24 O atoms. Thus, these transitions would have origin in these Sr-O interactions. The features localized between 16190 and 16200 eV exhibits only a little small decrease in the theoretical spectra (not shown) indicating that this transition is originated in the first Sr-Sr interactions. This information is consistent with the experimental XANES spectra with the increase of the incorporated Gd atoms, which substitute Sr atoms in the $SrTiO_3$ lattice.

Quantitative information about the local structure around a specific atom such as number of coordination, position, the type of the neighbor atoms and the disorder within the coordination shells can be obtained with EXAFS analysis. The experimental EXAFS spectra for our set of samples were obtained with the back Fourier transform (FT) between 1.0 and 5.5 Å. The theoretical model used in the fits were generated from FEFF9 code and structural data considering $SrTiO_3$ structure [35]. Fig. 2(a) and (b) shows the fitted $k\chi(k)$ and $k^3$ weighted FT of the SGTO samples, respectively. In these fits, the theoretical model, according to $SrTiO_3$ structure, consisted of single scattering paths corresponding to the five shells around the Sr absorber atom (From the first to the last shell: 12 O, 8 Ti, 6 Sr, 24 O and 11 Sr atoms). In the fits, the number of independent points is defined as $N_{ind} = 2\Delta R\Delta k/\pi$, where $\Delta R$ is the width of the $R$-space filter windows and $\Delta k$ is the actual interval of the fit in the $k$-space [36,37]. In all fits, $N_{ind}$ was kept greater than the number of free parameters ($N_{par}$). Table 1 exhibits the results of the best fits: interatomic distances ($R$), coordination number ($N$), and the parameters related to the reliability of the fit, the Debye-Waller factor ($\sigma^2$) and the quality factor ($QF$) [36,37]. Here, the total number of neighbors for each shell was kept fixed with the intention of minimizing $N_{par}$. This method was performed in order to prevent fitting drawbacks due to the signal/noise relation of the spectra ($N_{ind}$ = 11) on the parameter correlations.

In Fig. 2(b), the region between 1.0 and 2.5 Å in the FT is related to the single scattering paths of Sr-O interactions. The other single scattering paths associated with Sr-Ti, Sr-Sr and Sr-O (beyond the first O neighbors) interactions correspond to the peaks and shoulders identified beyond 2.5 Å. This region also includes multiple scattering paths such as Sr-O-O, Sr-O-Ti-O, Sr-O-O-Sr, Sr-Sr-O, Sr-Sr-Ti, Sr-O-O-O,

Sr-Ti-O, Sr-Ti-O-Ti and Sr-Ti-O-O interactions. The reliability of the fits is asserted by the values of the $\sigma^2$ and *QF* factors (Table 1). Nevertheless, the comparison of the fitted and experimental data in Fig. 2(a) and (b) also confirms the reliability of the fits. The values of distance between each shell and Sr absorber atom do not show changes within the uncertainty. On the other hand, Debye-Waller values show an increase as a function of the Gd content, mainly for Sr-O and Sr-Sr interactions. This variation is related to the substitutional disorder because of the difference of the crystal radius between $Gd^{3+}$ (1.41 Å) and $Sr^{2+}$ (1.58 Å) ions. The heterovalent nature of the $Gd^{3+}$ doping is also an expected source of disorder via induction of different kinds of point structural defects. Specifically, $Gd^{3+}$ ion substitutes $Sr^{2+}$ ion and 1/2 strontium vacancy ($V_{Sr}$) is formed via charge compensation effect. These vacancies are believed to break the translational symmetry of the lattice and represent a type of disorder associated with the formation of polar regions in the material [23]. Another mechanism of charge compensation is the formation of oxygen interstitial ($O_i$) when oxygen vacancies are presented. In this case, $Gd^{3+}$ ion substitutes $Sr^{2+}$ ion and 1/2 $O_i$ is formed [23]. Notably, the polymeric precursor method is a preparation process of oxides in which oxygen vacancies are common formed as a result of the synthesis and subsequent heat treatment [38]. Moreover, a decrease in the electrical conductivity due to a decrease in the number of oxygen vacancies because of the $Gd^{3+}$ incorporation has been reported [23].

EXAFS spectra were also collected at Gd $L_{III}$-edge for SGTO samples. Fig. 3(a) and (b) shows the fitted $k\chi(k)$ and $k^3$ weighted FT and the results obtained with the best fits are presented in Table 2. In this analysis, the experimental EXAFS spectra were obtained with the back FT between 1.5 and 2.5 Å. The theoretical model used in the fits were generated from FEFF9 code and crystallographic information considering $SrTiO_3$ where Gd absorber atom is incorporated at Sr sites. Although $GdTiO_3$ has a highly distorted perovskite structure with *Pbnm* space group [39], the fit considering Gd absorber atom at Sr sites, which means the Gd absorber atom surrounded by 12 O atoms at the same distance, is more reliable than the model with three shells with two O in the orthorhombic symmetry. As can be seen in Table 2, the parameters of the fit, within the uncertainties, does not changes substantially among the set of studied samples. We call attention to the obtained values of distance *R*, it is considerably smaller as compared to values of *R* for the first coordination shell (Sr-O) presented in Table 1 for the Sr, consistent to its smaller crystal radius as compared to that for the Sr.

The region beyond 2.5 Å in the Fig. 3(b) were also back FT in order to fit the resulting EXAFS spectra. However, the fits did not converge neither using shells according to the crystallographic cubic structure of the $SrTiO_3$ nor the orthorhombic structure of the $GdTiO_3$. Moreover, a decrease in the intensity of this region as a function of the Gd concentration is also observed in this figure. It is well known that the intensity of the FT is directly related to the coordination number and the Debye-Waller factor. Actually, the peaks and shoulders observed beyond 2.5 Å are ascribed to single scattering paths relative to Gd-Ti, Gd-Sr and Sr-O (beyond the first O neighbors) besides multiple scattering interactions. Thus, the decrease in the intensity and the unreliability of the fitting are due to the disorder associated with the break of the translational symmetry and the decreasing of the coordination number because of the formation of $V_{Sr}$. In other words, this region is a sensitive indicator of $V_{Sr}$ since Gd-Sr interactions are involved. As a result, we can infer that the $V_{Sr}$ are closely located to the Gd atoms in the $SrTiO_3$ lattice. Once the polymeric precursor method is well-known by the compositional homogeneity in the synthesis of oxide materials [40,41], the SGTO samples present vacancies homogeneously dispersed throughout the lattice in molecular level. This is an important issue when thermal and electric conductivity properties are considered. Additionally, although it has been reported that $Gd^{3+}$ ions occupy off-center positions in the dodecahedral $SrO_{12}$ sites [24,25], our EXAFS analysis indicates regular $GdO_{12}$ dodecahedra without displacement of Gd atoms from centrosymmetric position.

Fig. 4 exhibits the XANES spectra at the Ti $L_{III,II}$-edges for the SGTO samples. A subtracting of the background was performed with two arctangent functions. Forbidden transitions in the L–S coupling (spin–orbit coupling), although permitted because of the *pd* multipolar interactions, are attribute to the origin of the peaks A and B [42]. The Ti $L_{III}$-edge comprises the features C and D, which correspond to the split in 3*d* band into $t_{2g}$ and $e_g$ subbands, respectively [43]. It is relevant to emphasize that the Ti 3*d* $e_g$ subband (D) still splits into $d_{x^2-y^2}$ and $d_{z^2}$ orbitals that point to the four side-corners and the two apex O ions of the Ti octahedron, respectively [43,44]. This infers that the feature D is directly influenced by Ti off-center displacement due to the variation in the lengths of Ti–O bond. On the other hand, feature labeled as C is not affected by Ti off-center displacement because the $3d_{xy}$, $3d_{yz}$, and $3d_{zx}$ orbitals of the $t_{2g}$ subband point in directions between the O ions [43]. The Ti $L_{II}$-edge includes the features labeled as E and F, which have also origins in the split of the 3*d* band into $t_{2g}$ and $e_g$ subbands caused by the crystal field. Similarly, the split of the $e_g$ states in the F peak also arises, although this is not well-resolved because of the lifetime-related broadening [43].

As can be also seen in Figs. 4 (yellow highlighted area), the peak labeled as D for STO sample is evidently asymmetric and cannot be fitted using just one peak. This statement is consistent with the splitting of the $e_g$ orbitals, as commented before, and indicates a Ti displacement relative to its centrosymmetric position in TiO$_6$ octahedra. This displacement is in agreement with STO composition synthesized by the polymeric precursor method whose static displacement of about 0.04 Å in the [001] direction was evaluated through Ti K-edge XANES spectra [35]. It is also observed in Fig. 4 that the labeled peak D becomes larger with the incorporation of Gd atoms, demonstrating a more discernible splitting of $e_g$ orbital. Therefore, this enlargement denotes a Ti off-center displacement and a small deviation from cubic structure due the local symmetry breaking associated with Gd substitution. Only in high symmetric TiO$_6$ octahedra the Ti 3$d$ can form triply energy degenerated $t_{2g}$ states (3$d_{xy}$, 3$d_{yz}$, and 3$d_{zx}$) leading to a large DOS close the bottom of the conduction band (the sum of the DOS correspondent to the three states). In distorted TiO$_6$ octahedra the energy degeneracy of the Ti 3$d$ orbitals is broken with the 3$d_{xy}$ state lying alone at the bottom of the conduction band, leading to a relatively smaller DOS [45]. As a consequence, a smaller DOS results in higher values of the effective mass of carrier electrons, which would led to higher values for the Seebeck coefficient [46]. And, in fact, one can find in the literature reports of large absolute values of the Seebeck coefficient upon Gd doping [21,47,48]. Similar behavior has been also observed for PbTiO$_3$ based systems, where homovalent or heterovalent substitutions of Pb atoms by Ca, Ba or La atoms induces the decreasing in the $c/a$ tetragonal ratio and in the Ti off-center displacement [26,49,50]. As a consequence, the peak labeled as D becomes narrower with the addition of the incorporated atoms [26,49,50]. Otherwise, the peak labeled as D could be adjusted by a simple Gaussian function for XANES spectra of Ba$_{1-x}$Sr$_x$TiO$_3$ system, implying that there is no splitting in $e_g$ orbitals, corresponding to a negligible tetragonal distortion and/or a lower $c/a$ ratio, indicating that the centrosymmetric positions of the TiO$_6$ octahedra are occupied by Ti atoms [51].

Figure 5(a) presents the XANES spectra at O K-edge for SGTO samples. The main spectral features are centered at ~ 534, 537, 539, 542, and 547 eV (labeled as A, B, C, D and E, respectively), corresponding to transitions from the O 1$s$ core state to the unoccupied O 2$p$-derived states [51]. The inset of Fig. 5(a) shows a spectral broadening and a decrease in intensity of the features A, B, C and D as a function of Gd doping. The spectral broadening can be viewed as a signal of the growing hybridization between the O 2$p$ and Gd states and the decrease of the intensities indicates a decrease of the average unoccupied O 2$p$-derived states with the Gd doping. Theoretical XANES spectrum for a cluster with a radius of 6.0 Å was calculated at the O K-edge of SrTiO$_3$ by using FEFF9 code. Figure 5(b) exhibits the calculated results and an illustration of the cluster used in the calculation. We observe that the theoretical spectrum reasonably reproduces the features observed experimentally A, B, C, D and E. It was reported that O 2$p$-projected DOS is similar to the experimental XANES spectra at O K-edge for other perovskite systems [26,52,53]. Then, under equivalent conditions to the calculation of the theoretical XANES spectrum, *ab initio* quantum-theoretical calculations of the local density of states (these calculations involve $s$-, $p$-, $d$- and $f$-projected DOS) of the oxygen, titanium, strontium and gadolinium atoms were performed. Figure 6 shows the results of the local DOS calculation, the presented energy scale is as calculated by the FEFF9 code. We observe that the position of the maxima of Ti 3$d(t_{2g})$, Sr 4$d$ and Gd 4$f$5$d$ [54] calculated DOS coincides with the first peak in the O 2$p$ DOS (between −11.0 and −9.0 eV), which signifies that transition centered at ~534 eV (A) results mainly from the transitions of the O 1$s$ orbital to antibonding O 2$p$ states hybridized with Ti 3$d(t_{2g})$ (major contribution), Sr 4$d$ and Gd 4$f$ 5$d$ (minor contributions) orbitals. It is important to note in Fig. 6 that the Ti 3$d$ DOS intensity is divided by 13. As pointed earlier, this feature in the SGTO experimental spectra is dependent on the Gd content (inset of Fig. 5(a)). As can also be seen in Fig. 6, the region between −9 and −6 eV in the O 2$p$ DOS coincides with Ti 3$d(e_g)$, Sr 4$d$ (main contribution) and Gd 4$f$5$d$ calculated DOS. Therefore, the O 2$p$ states hybridized with Ti 3$d(e_g)$, Sr 4$d$ and Gd 4$f$5$d$ orbitals can be associated with the features in the XANES spectra between 537 and 539 eV (B and C), which also exhibit a dependence of the Gd content. Following this line of reasoning, we infer that the absorption peak centered at 542 eV (D) arises from the hybridization of O 2$p$ with Sr 4$d$ and, also, with Gd 4$f$5$d$ (between -6 and -4 eV), while the peak at 547 eV (E) correspond to the hybridization of O 2$p$ with Ti 4$pf$ (between -2 and 4 eV).

According to this analysis, the decrease of the intensity of the features A, B, C and D (inset of Fig. 5(a)) can be understood in terms of the difference between electronegativity values of Sr (0.95) and Gd (1.20) and the relative small DOS for the Gd (Fig. 6(d)) as compared mainly to the DOS for the Ti 3$d$ and Sr 4$d$. With the Gd doping the average electronegativity of the Sr sites of the SrTiO$_3$ structure increases, what leads to an increase of the hybridization between the O 2$p$ and Gd 4$f$5$d$ states, in detrimental of the Ti 3$d$($t_{2g}$) for the feature A and Ti 3$d$($e_g$) in the case of the feature B and C [26,51]. Since the DOS for the Gd 4$f$5$d$ is relatively lower in respect to the DOS for the Ti 3$d$ state, the increase of the hybridization O 2$p$-Gd 4$f$5$d$ leads to a decrease of the number of accessible states (unoccupied states) and the experimental observed decrease in the intensity of the features A, B and C. The same argumentation also applies for the case of feature D, associated to the hybridized states O 2$p$, Sr 4$d$ and Gd 4$f$5$d$. Furthermore, Ti 3$d$ and 4$s$ states contributes to the conduction band with little composition of O 2$p$ and Sr 5$s$ states whereas the valence band is formed manly by O 2$p$ states with minor constituent of Ti 4$s$ and 3$d$ states [46,55]. According to our DOS calculation, we estimate the SGTO band gap ($E_g$) from the difference in energy between the two highest peaks in the DOS for the Ti 3$d$ (-9.7 eV) and O 2$p$ (-6.5 eV) states, which gives $E_g \approx 3.25$ eV, bigger than some previous theoretical calculations but in agreement with experimental value of 3.22 eV [2,39,55]. Additionally, Fig. 6 shows that DOS for the Gd 4$f$ states coincide with the position of the DOS for the Ti 3$d$($t_{2g}$) (between −11.0 and −9.0 eV) giving a contribution to the conduction band. Such hybridization, besides the structural distortion mentioned early, can explain why the conductivity and mainly the Seebeck coefficient is considerably affected by the Gd and, more generally, by the STO rare-earth doping [56, 57].

## 4. Conclusions

In this study, local and electronic structure of SGTO samples were analyzed through XAS measurements. The results obtained with the adjustment of EXAFS spectra at Sr K-edge show that crystallographic model of *Pm-3m* space group is consistent with local structure around Sr atoms, as expected. This same analysis also reveals an increasing of the Debye-Waller as a function of the Gd content some shells, which is associated with disorder induced by Sr vacancies due to the heterovalent Gd incorporation. EXAFS spectra at Gd L$_{III}$-edge for SGTO samples indicates regular GdO$_{12}$ dodecahedra without displacement of Gd atoms from centrosymmetric position. A disorder was also identified in the shells beyond the first 12 O neighbors in which neither the crystallographic cubic structure of the SrTiO$_3$ nor the orthorhombic structure of the GdTiO$_3$ fits well. XANES spectrum at Ti L$_{III,II}$-edges shows an asymmetric peak because of the splitting between the $e_g$ orbitals of 3$d$ band for STO sample. The addition of Gd atoms to SrTiO$_3$ structure cause an enlargement of this peak and this split is associated with a small displacement of Ti atoms from their centrosymmetric position. Several features of the XANES spectra at O K-edge for SGTO samples are affected by the increase of Gd concentration. According to our calculated projected density of states, these transitions are related to a reduction in the number of unoccupied O 2*p*-derived states.

### 5. Acknowledgements

This research used resources of the Brazilian Synchrotron Light Laboratory (LNLS), an open national facility operated by the Brazilian Centre for Research in Energy and Materials (CNPEM) for the Brazilian Ministry for Science, Technology, Innovations and Communications (MCTIC) (Proposal numbers 20180277 and 20180459). The XAFS2 and PGM beamline staffs are acknowledged for the assistance during the experiments.

# TABLES

**Table 1** - Sr K-edge EXAFS simulation results. $R$ is the distance from the absorber atom, $N$ is the average coordination number (not fitted), $\sigma^2$ the Debye-Waller factor and $QF$ the quality factor.

| Sample | Shell | $R$ (Å) | $N$ | $\sigma^2$ (Å$^2$) | $QF$ |
|---|---|---|---|---|---|
| STO | Sr-O | 2.71(1) | 12 | 0.012(2) | |
| | Sr-Ti | 3.39(1) | 8 | 0.006(1) | |
| | Sr-Sr | 3.92(2) | 6 | 0.010(2) | 1.54 |
| | Sr-O | 4.76(3) | 24 | 0.012(2) | |
| | Sr-Sr | 5.50(3) | 11 | 0.010(2) | |
| SGTO6 | Sr-O | 2.70(2) | 12 | 0.014(2) | |
| | Sr-Ti | 3.40(1) | 8 | 0.007(1) | |
| | Sr-Sr | 3.92(2) | 6 | 0.011(2) | 1.50 |
| | Sr-O | 4.76(5) | 24 | 0.014(2) | |
| | Sr-Sr | 5.49(5) | 11 | 0.011(2) | |
| SGTO12 | Sr-O | 2.68(2) | 12 | 0.016(2) | |
| | Sr-Ti | 3.39(1) | 8 | 0.006(1) | |
| | Sr-Sr | 3.92(3) | 6 | 0.011(1) | 1.15 |
| | Sr-O | 4.75(5) | 24 | 0.016(2) | |
| | Sr-Sr | 5.473(5) | 11 | 0.011(1) | |
| SGTO18 | Sr-O | 2.67(2) | 12 | 0.017(3) | |
| | Sr-Ti | 3.40(1) | 8 | 0.007(1) | |
| | Sr-Sr | 3.92(2) | 6 | 0.012(2) | 1.02 |
| | Sr-O | 4.74(3) | 24 | 0.017(3) | |
| | Sr-Sr | 5.53(5) | 11 | 0.012(2) | |
| SGTO24 | Sr-O | 2.67(2) | 12 | 0.018(3) | |
| | Sr-Ti | 3.40(1) | 8 | 0.007(1) | |
| | Sr-Sr | 3.90(3) | 6 | 0.012(2) | 1.09 |
| | Sr-O | 4.74(5) | 24 | 0.018(3) | |
| | Sr-Sr | 5.53(5) | 11 | 0.012(2) | |
| SGTO30 | Sr-O | 2.66(2) | 12 | 0.018(2) | |
| | Sr-Ti | 3.39(1) | 8 | 0.007(1) | |
| | Sr-Sr | 3.90(2) | 6 | 0.012(1) | 0.85 |
| | Sr-O | 4.73(4) | 24 | 0.018(3) | |
| | Sr-Sr | 5.51(5) | 11 | 0.012(1) | |

**Table 2 -** Gd L$_{III}$-edge EXAFS simulation results. *R* is the distance from the absorber atom, *N* is the average coordination number, $\sigma^2$ the Debye-Waller factor and *QF* the quality factor.

| Sample | Shell | *R* (Å) | *N* | $\sigma^2$ (Å$^2$) | *QF* |
|---|---|---|---|---|---|
| SGTO6  | Gd-O | 2.43(3) | 9(2) | 0.022(5) | 2.17 |
| SGTO12 | Gd-O | 2.39(1) | 8(1) | 0.019(4) | 0.62 |
| SGTO18 | Gd-O | 2.38(1) | 9(1) | 0.019(4) | 0.47 |
| SGTO24 | Gd-O | 2.37(1) | 9(1) | 0.018(3) | 0.83 |
| SGTO30 | Gd-O | 2.37(1) | 9(1) | 0.016(3) | 0.48 |

# FIGURE CAPTIONS

**Figure 1** – (a) XANES spectra at Sr K-edge for the SGTO samples. (b) Theoretical XANES spectra at Sr K-edge as a function of the cluster radius used in the calculation. The clusters are also illustrated; the pink, red, grey and green spheres represent the absorber Sr, O, Ti and non-absorber Sr atoms, respectively. The insets in (a) and (b) highlight the main differences between the not offset spectra.

**Figure 2** – (a) $k\chi(k)$ spectra and (b) corresponding $k^3$ weighted Fourier transforms (FT) of the X-ray absorption spectra obtained at Sr K-edge for the SGTO samples. Open symbols are the experimental data and the solid lines represent the fittings using the parameters shown in Table 1. The spectra are offset for clarity.

**Figure 3** – (a) $k\chi(k)$ spectra and (b) corresponding $k^3$ weighted Fourier transforms (FT) of the X-ray absorption spectra obtained at Gd $L_{III}$-edge for the SGTO samples. Open symbols are the experimental data and the solid lines represent the fittings using the parameters shown in Table 2. The spectra are offset for clarity.

**Figure 4** – XANES spectra at Ti $L_{III,II}$-edge for the SGTO samples.

**Figure 5** – (a) XANES spectra at O K-edge for the SGTO samples. (b) Calculated XANES spectra for the undoped STO sample. The cluster used in the calculation is also illustrated in (b); here the pink, red, grey, and green spheres represent the absorber O, the non-absorber O, Ti, and Sr atoms, respectively.

**Figure 6** – Projected DOS of the constituent elements obtained via FEFF9 code. The dotted line DOS correspond to states not involved into the calculation of the O K-edge XANES spectra (Fig. 5(b)).

**FIGURES**

*Figure* 1

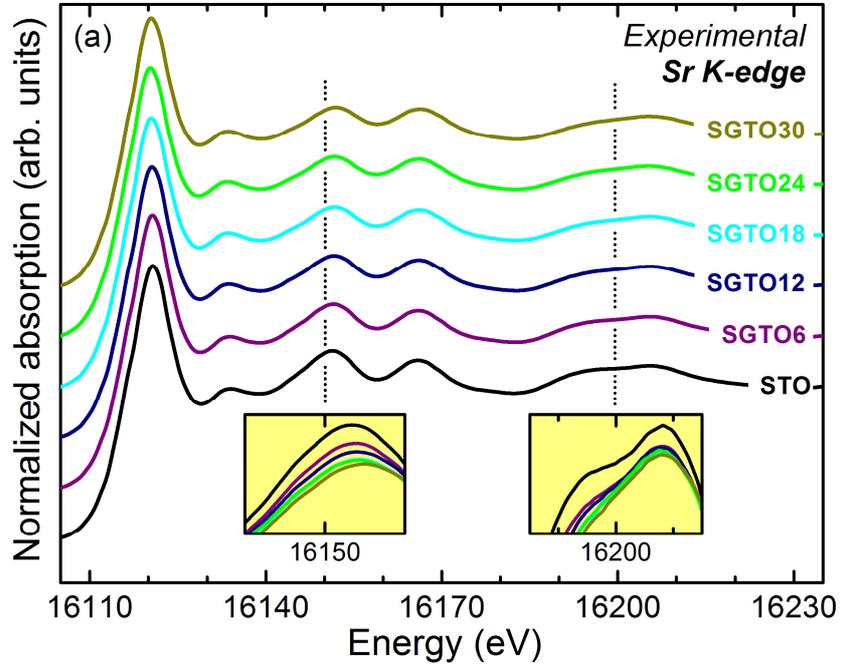

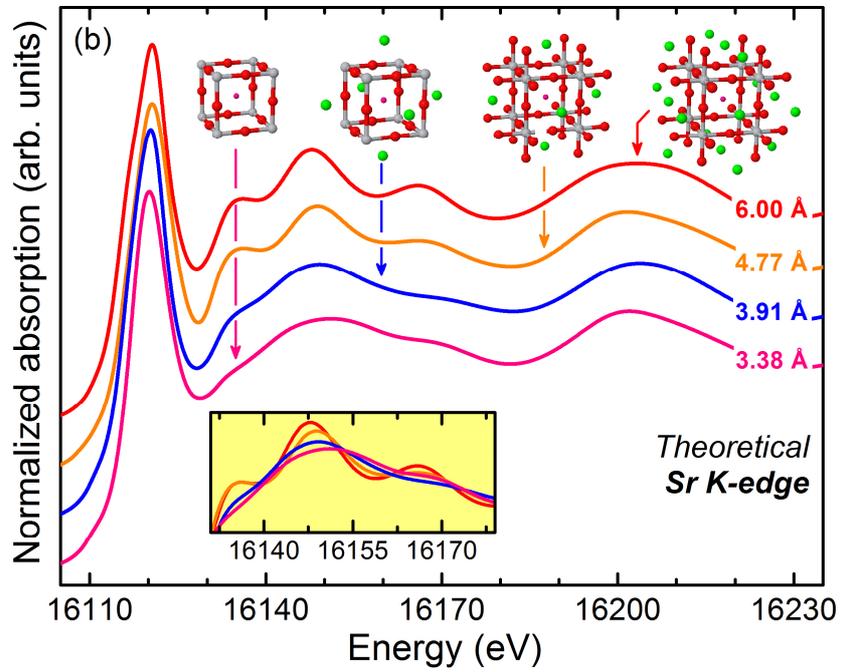

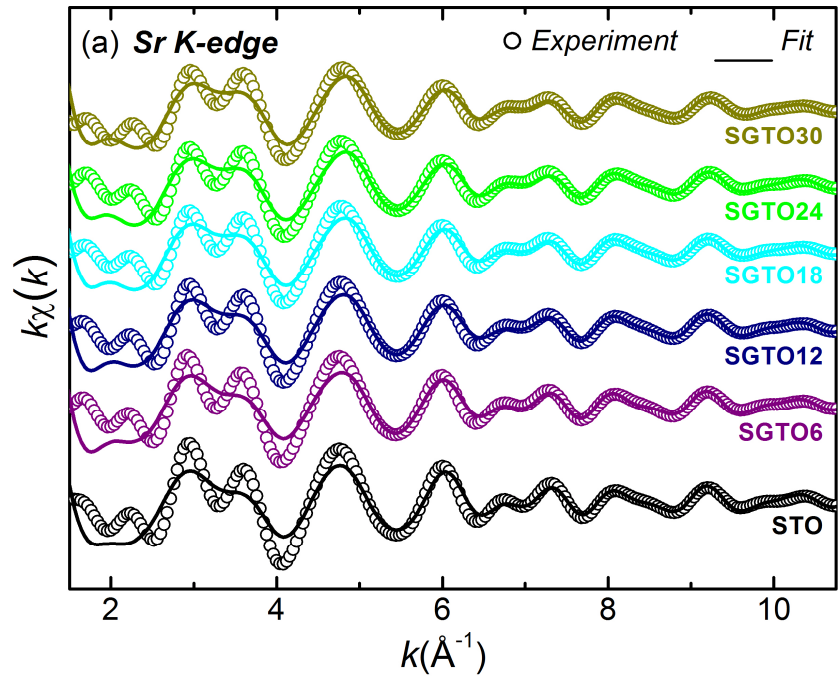

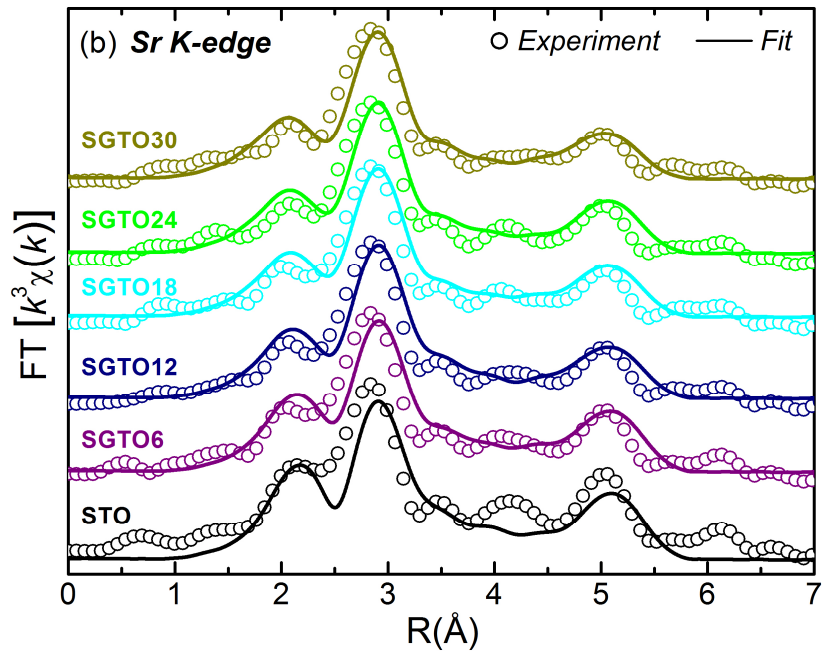

*Figure 3*

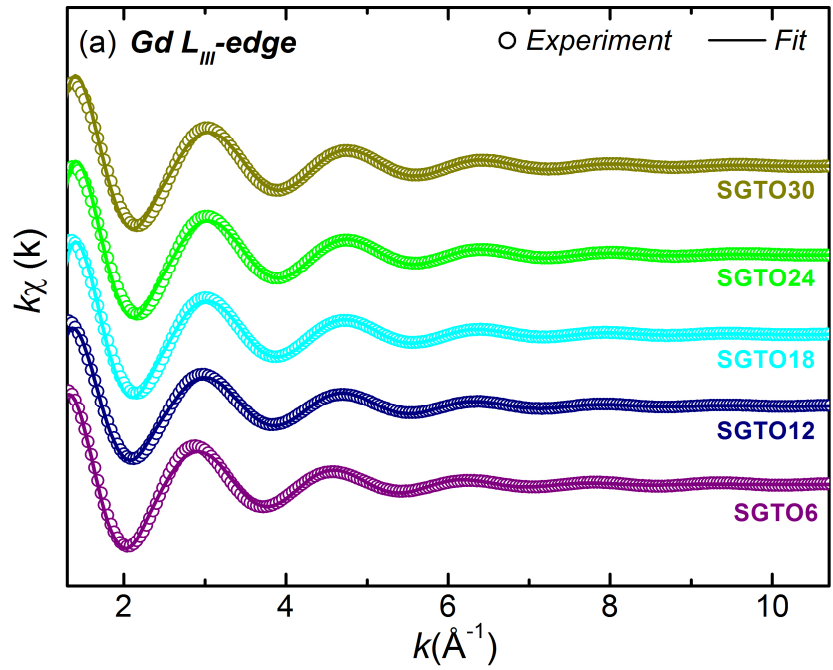

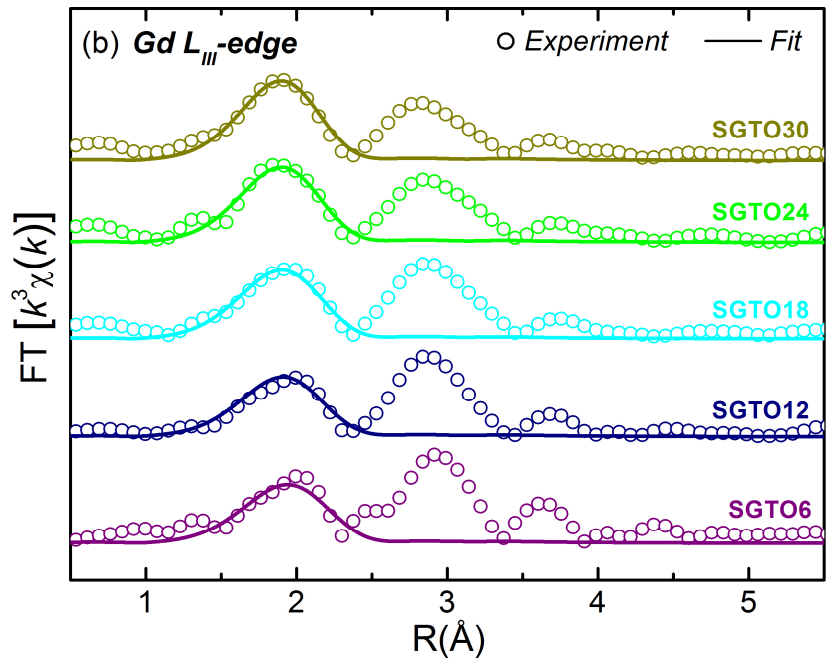

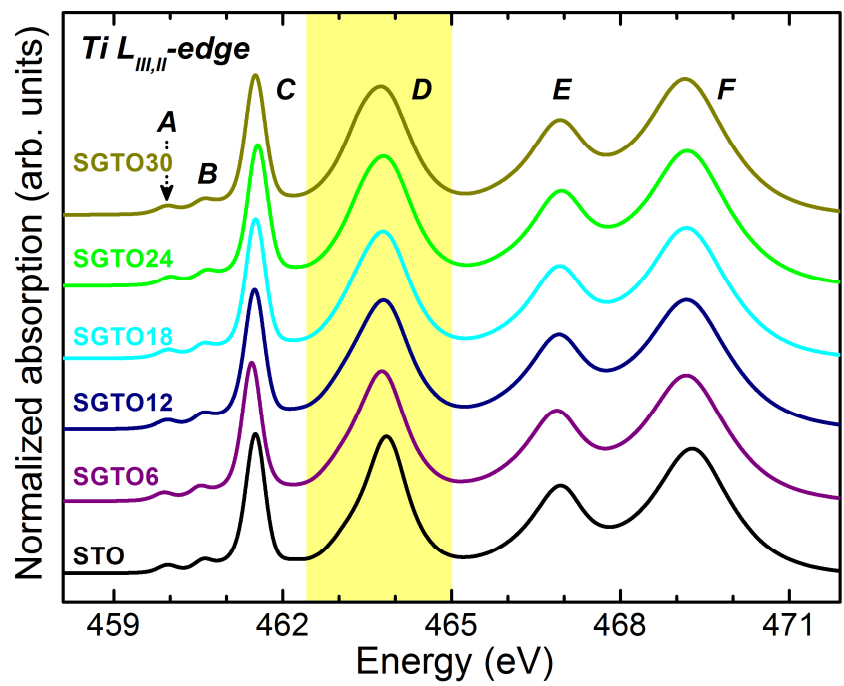

*Figure 5*

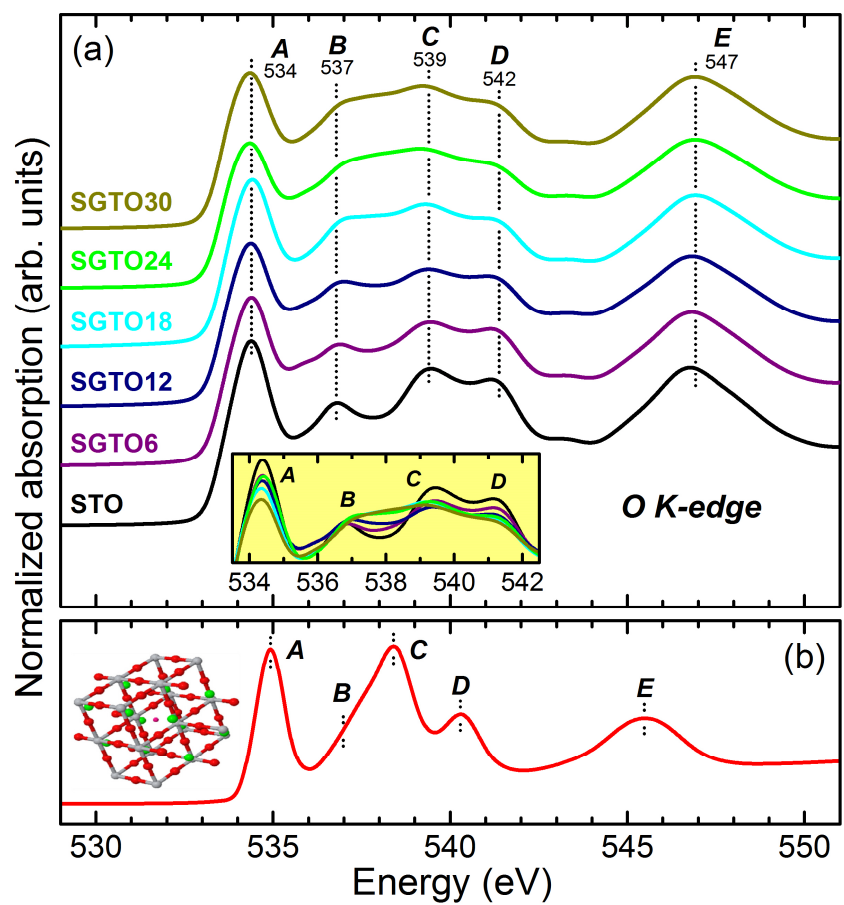

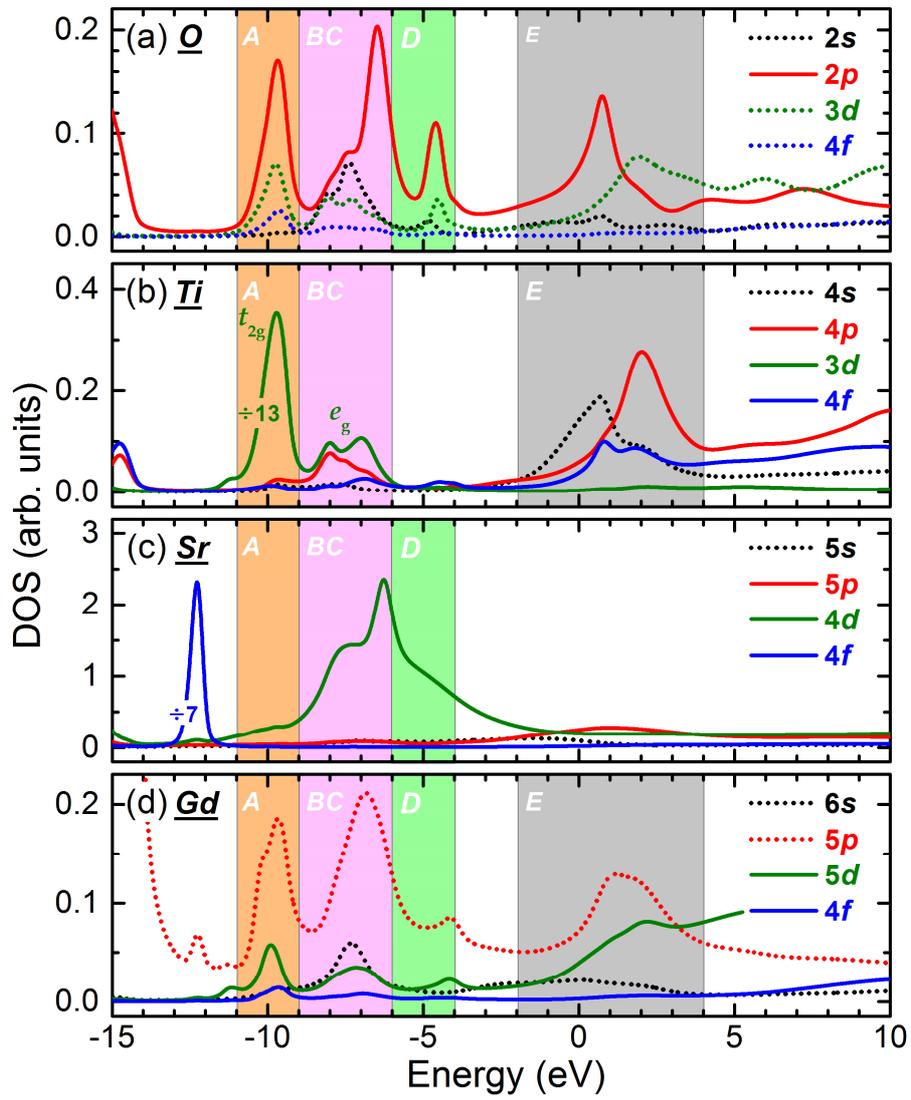